\documentclass[a4paper,fleqn,usenatbib]{mnras}

\usepackage{newtxtext,newtxmath}

\usepackage[T1]{fontenc}
\usepackage{ae,aecompl}

\usepackage{graphicx}	
\usepackage{amsmath}	
\usepackage{amssymb}	

\usepackage[linesnumbered,boxed,lined,commentsnumbered]{algorithm2e}
\usepackage{tikz}

\usepackage{relsize}
\graphicspath{{Figs/}}

\usepackage{soul}

\title[Centroid bias in Shack-Hartmann wavefront sensing]{Peak-locking centroid bias in Shack-Hartmann wavefront sensing}

\author[N. Anugu et al.]{
Narsireddy Anugu,$^{1,2}$\thanks{E-mail: narsireddy.anugu@fe.up.pt}
Paulo J. V. Garcia,$^{1}$
Carlos M. Correia$^{3}$
\\
$^{1}$Faculdade de Engenharia, Universidade do Porto, rua Dr. Roberto Frias, 4200-465 Porto, Portugal;\\ 
CENTRA -- Centro de Astrof\'{\i}sica e Gravita\c c\~{a}o, IST, Universidade de Lisboa, P-1049-001 Lisboa, Portugal\\
$^{2}$School of Physics, Astrophysics Group, University of Exeter, Stocker Road, Exeter EX4 4QL, UK\\
$^{3}$Aix Marseille Univ, CNRS, LAM, Laboratoire d'Astrophysique de Marseille, Marseille, France.
}

\date{Accepted XXX. Received YYY; in original form ZZZ}

\pubyear{2015}

\begin{document}
\label{firstpage}
\pagerange{\pageref{firstpage}--\pageref{lastpage}}
\maketitle

\begin{abstract}
Shack-Hartmann wavefront sensing relies on accurate spot centre measurement. Several algorithms were developed with this aim, mostly focused on precision, i.e. minimizing random errors. In the solar and extended scene community, the importance of the accuracy (bias error due to peak-locking, quantisation or sampling) of the centroid determination was identified and solutions proposed. But these solutions only allow partial bias corrections. To date, no systematic study of the bias error was conducted. This article bridges the gap by quantifying the bias error for different correlation peak-finding algorithms and types of sub-aperture images and by proposing a practical solution to minimize its effects. Four classes of sub-aperture images (point source, elongated laser guide star, crowded field and solar extended scene) together with five types of peak-finding algorithms (1D parabola, the centre of gravity, Gaussian, 2D quadratic polynomial and pyramid) are considered, in a variety of signal-to-noise conditions. The best performing peak-finding algorithm depends on the sub-aperture image type, but none is satisfactory to both bias and random errors. A practical solution is proposed that relies on the anti-symmetric response of the bias to the sub-pixel position of the true centre. The solution decreases the bias by a factor of $\sim 7$ to values of $\lesssim 0.02~\mathrm{pix}$. The computational cost is typically twice of current cross-correlation algorithms.
\end{abstract}

\begin{keywords}
Instrumentation: adaptive optics -- Techniques: high angular resolution -- Techniques: image processing
\end{keywords}



\section{Introduction}\label{sec:intro}

The Shack-Hartmann wavefront sensor is commonly used to measure the wavefront aberrations in astronomical adaptive optics \citep{Tyson2015}, optical testing \citep{Malacara2007}, ophthalmology \citep{Burns2014} or microscopy \citep{Booth2014}. It consists of a two dimensional (2D) array of micro lenses. For a plane wavefront incidence, the spots are focused on the optical axis of the each micro lens -- the reference centres. For an aberrated wavefront, the imaged spots are displaced from the reference centres. The estimation  of the spot displacements between the aberrated  and the reference spots allows one to retrieve the incident aberrated wavefront profile \citep{Dai1996} 

Correlation algorithms are used to estimate spot displacements when extended sources are present (cf. \cite{Rais2016} for a recent review). The cross-correlation: a) is optimal at lower signal-to-noise ratios \citep{VijayaKumar1992} and; b) is fast and of simple implementation over other methods such as maximum likelihood \citep{Gratadour2005} and iterative gradient-based shift estimators \citep{Rais2016}; c) has unitary gain \citep{Gratadour2010}. Cross-correlation is applied to measure image displacements for solar adaptive optics \citep{Woger2009, Lofdahl2010,  Townson2015}, laser guide star elongated spots \citep{Thomas2008, Basden2014} and to extended scene wavefront sensing \citep{Poyneer2003, Robert2012}.  The image displacement  is computed  by cross-correlating a reference\footnote{Cf. \cite{Basden2014} for several approaches for reference image generation.} image to the target aberrated sub-aperture image.  The  correlation algorithm can be implemented  either in spatial domain \citep{Lofdahl2010} or  in the Fourier domain \citep{Poyneer2003, Sidick2013}. In both domains, the image displacement is measured in two steps.  In the first step, the cross-correlation  between the reference  and the target image is computed. In the second  step, a sub-pixel peak-finding algorithm is applied to the correlation image  \citep{Poyneer2003}. Commonly used  peak-finding algorithms in image registration are 1D parabola fitting \citep{Poyneer2003, Thomas2006, Robert2012},  Gauss fitting \citep{Nobach2005}, centre of gravity, pyramid fitting \citep{Bailey2003} and 2D quadratic polynomial fitting \citep{Lofdahl2010}. These will be addressed further in the article (cf. Table~\ref{tab:centroid_alg}).

\begin{table*}
	\caption{\label{tab:centroid_alg} Sub-pixel ($s'_x$, $s'_y$)  peak-finding algorithms. The estimate of $s'_y$  is obtained in an analogous fashion to $s'_x$. }
	\vspace{5pt}
	
	\begin{tabular}{l}
		\hline Algorithm \\\hline\\
		
		1D Parabola fit (PF):\\
		
		$s'_x = x_0 + 0.5 \times \dfrac{  C[x_0-1, y_0] - C[x_0+1, y_0]  }{  C[x_0-1, y_0] + C[x_0+1, y_0]  - 2 C[x_0, y_0]  } $ \\\\
		
		Gaussian fit (GF):\\
		
		$s'_x = x_0 + 0.5 \times \dfrac{ \ln(C[x_0-1, y_0]) -  \ln(C[x_0+1, y_0])  }{  \ln(C[x_0-1, y_0]) + \ln(C[x_0+1,y_0])  - 2 \ln(C[x_0,y_0])  }$ \\\\
		
		Pyramid  (PYF):\\
		$s'_x = x_0 + 0.5 \times \dfrac{C[x_0-1, y_0] - C[x_0+1, y_0]}{ \min\big( C[x_0-1, y_0], C[x_0+1, ~y_0] \big) -  C[x_0, y_0] }$  \\\\
		
		2D Quadratic polynomial fit (QPF)\\[6pt]
		($s'_x$, $s'_y$) $= \Big( x_0 + \dfrac{2a_1a_5-a_2a_4}{a_4^2 - 4a_3a_5}, y_0 + \dfrac{2a_2a_3-a_1a_4}{a_4^2 - 4a_3a_5} \Big) $\\
		with  polynomial $f(x)=a_0 + a_1 x + a_2 y+a_3 x^2 + a_4 x y + a_5 y^2$. \\\\
		
		Centre of Gravity  (CoG)\\ 
		$s'_x = x_0 +  \dfrac{C[x_0-1, y_0] - C[x_0+1, y_0]}{ 3 \min\big( C[x_0-1, y_0], C[x_0+1, y_0] \big) - \big( C[x_0, y_0] + C[x_0+1, y_0] + C[x_0-1, y_0] \big)}$  \\\\
		\hline
	\end{tabular}
\end{table*}

Sub-pixel peak-finding in the correlation image is biased towards integer pixels. In adaptive optics, these errors are often referred as systematic bias errors, quantisation errors or sampling errors. Methods for their correction are modelling and a posteriori correction  \citep{Woger2009, Lofdahl2010, Sidick2013}. These approaches are limited because the   bias errors depend on: a) modelling; b) the sub-aperture image characteristics; c) the noise level; d) the combination of correlation and the peak-finding algorithms; making it difficult to model -- especially in low signal-to-noise conditions.

In the following, the bias problem of centroid algorithms is addressed. In Sec.~\ref{sec:methods} the methods used are presented, including a novel algorithm for bias error reduction. The results on the bias performance of several peak-finding algorithms are presented in Sec.~\ref{centreResults}. It is found that no algorithm is simultaneously satisfactory for both bias and random errors.  The results on the proposed solution to the bias error are presented and discussed in Sec.~\ref{TwoStepResults1}. In Sec.~\ref{sec:conclusion} we conclude by recalling the main ideas.

\section{Methods}\label{sec:methods}

\subsection{Current peak-finding methods}\label{sec:existing_alg} 

Consider  the reference ($I_0$) and sub-aperture ($I_\mathrm{S}$) images, with size $N\times N\,\mathrm{pix}^2$.  The cross-correlation in the image domain ($C$),  is given by 

\begin{equation}\label{CCI}
C[m,n] = \sum_{i=1}^{N} \sum_{j=1}^{N} I_\mathrm{S}[i + m, j + n]~I_0[i,j],
\end{equation}

\noindent The image displacement in integer pixels is determined from the correlation maximum location, which is at the pixel $(x_0, y_0)$. The sub-pixel image displacement ($s'_x$,  $s'_y$) is estimated by applying  2D centroid algorithms (cf. Table~\ref{tab:centroid_alg} and Sec.~\ref{sec:intro}) to the correlation  map $C[m,n]$. In most algorithms only five pixels are used: $(x_0, y_0)$, $(x_0-1, y_0)$ , $(x_0+1, y_0)$ , $(x_0, y_0-1)$ and $(x_0, y_0+1)$. For the 2D quadratic polynomial fit nine pixels instead of five are required for the estimation of the six coefficients. The estimation of $s'_y$ is analogous to $s'_x$.

\begin{figure}
	\centering
	\parbox[t]{11pt}{\rotatebox{90}{\hspace{0.3\columnwidth}  $\beta$ (pix)}}
	\includegraphics[width=0.7\columnwidth]{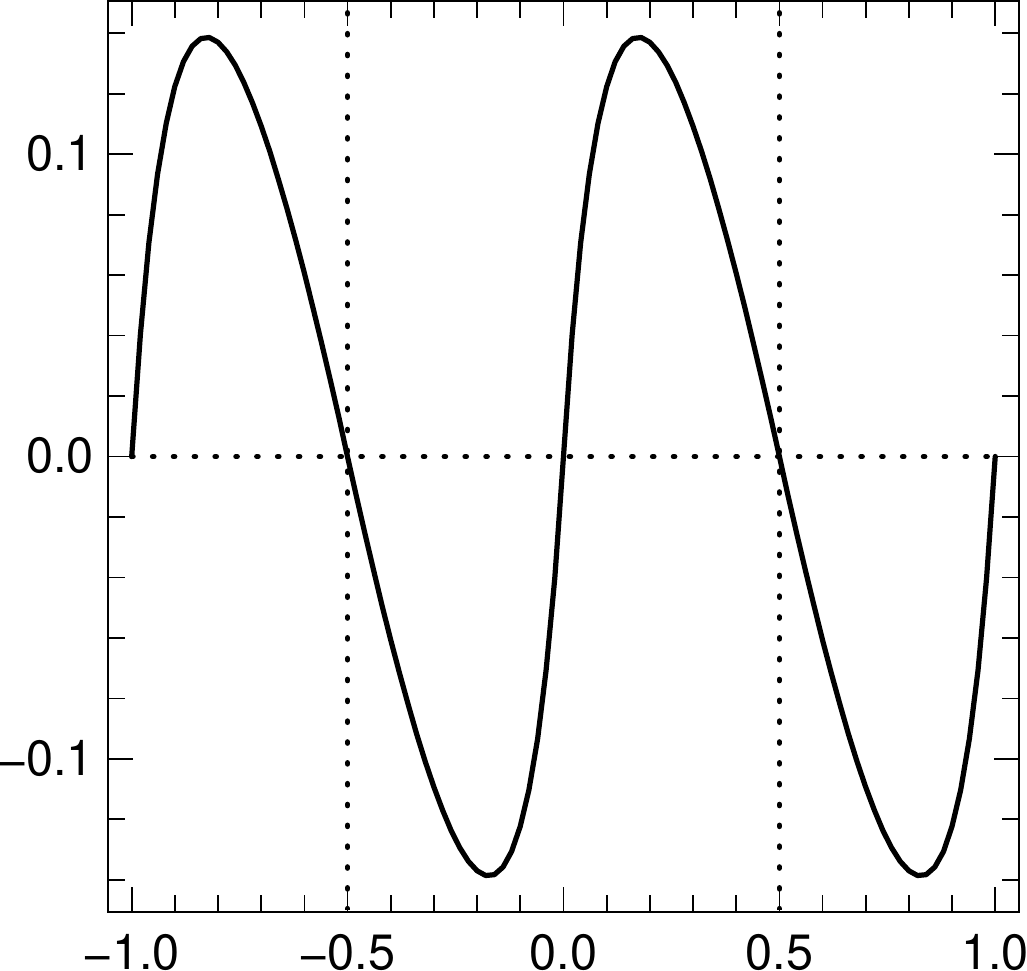}\\
	\hspace{16pt} $s$ (pix)
	\caption{\label{fig:BiasError} Bias error for a point source with centre of gravity centroid algorithm versus $s_x$. The shift vector is $\vec{s}=[s,s]^T$, i.e.  $s_x=s$. }
\end{figure}

The measured  displacement $s'$ (in a given direction $x$ or $y$) by the centroid algorithms is related to the real displacement $s$ by
\begin{equation}\label{eq:error}
s'=s+\beta+\epsilon,
\end{equation}
where $\beta$ is the bias error and $\epsilon$ the noise error. As referred in the Introduction these algorithms have systematic errors, the bias error $\beta$ exhibits a "sinusoidal" variation with an exact shape depending on $\epsilon$, image and centroid algorithm. In Fig.~\ref{fig:BiasError} an example of this bias is presented, in the absence of noise, for the cross-correlation algorithm (Eq. \ref{CCI}).  The origin of the bias is well known in the strain measurement community, it is due to the transfer function of the centroid algorithm \citep[e.g.][]{Schreier2000}. For example, the transfer function of the linear interpolation is not unitary but a complex number. Its module and phase changes with interpolation position. Therefore a bias in intensity and shift when an interpolation is made \citep[cf.][for details]{Schreier2000}.

In the presence of noise $\epsilon$, the  bias error $\beta$ in Eq.~\ref{eq:error} is estimated by taking the average of a large number of realizations, assuming that $s$ is constant for the number of realizations. The noise error is then significantly  reduced and Eq.~\ref{eq:error} becomes

\begin{equation}\label{eq:BiasError}
\left<s'\right> \simeq s+\beta,
\end{equation}
\noindent where the $\left<\right>$ denotes average. 

The noise error $\epsilon$ is estimated by the root-mean-square (RMS) deviation of the random sample of realizations
\begin{equation}\label{SNR}
\epsilon \simeq \sigma =  \sqrt{\frac{\sum_{i=1}^n \left( s'-\left<s'\right> \right)^2}{n}},
\end{equation}
\noindent with $n$ the number of realizations.

\subsection{Window shift peak-finding algorithm}\label{Methods2} 

In the standard approach, the sub-pixel peak centre is determined by directly applying a peak-finding algorithm of Table~\ref{tab:centroid_alg}. In this work, a method to reduce the bias error in the peak-finding is proposed. A similar method was previously applied in the context of particle image velocimetry \citep{Gui2002}, but to our knowledge, it is presented for the first time in the context of adaptive optics.

It is a two-step method: a) coarse search; b) fine search.  In Algorithm~\ref{alg:Method2} the pseudo-code of the method is presented. In the first step (coarse search, lines 2 to 4) the integer pixel maximum location $(x_0,y_0)$ is found. In the second step, (fine search, lines 5 to 19) an image region of interest ($I_\mathrm{ROI}$, cf. line 12) is interpolated from the sub-aperture image $I_\mathrm{S}$. The interpolation is made with the same sampling as the original image. At each iteration, the interpolation is done at changing fractional initial positions $\delta$ (cf. line 7). Then the correlation between the reference $I_0$ and each $I_\mathrm{ROI}$ is computed (cf. line 13). The sub-pixel displacements $s'$ are then obtained using the peak-finding algorithms of Table~\ref{tab:centroid_alg} (cf. line 16). These sub-pixel displacements are then corrected by the step $\delta$ introduced during the interpolation (cf. line 17). This process is repeated $K$ times, with varying $\delta$ (cf. line 7).  Because $K$ correlations took place,  $s'$ is a vector of $K$ elements. The individual displacements $s'[k]$ are affected by the bias $\beta$. This bias is "sinusoidal" and anti-symmetric, with period 1\,pix, as referred in Fig.~\ref{fig:BiasError}. The algorithm then takes the average of all $K$ displacements (cf. line 19), which reduces the bias approximately proportional to $K$.

For computational efficiency the cross-correlation $C$ is not computed in all pixels but only for a sub-image of size $5\times5\,\mathrm{pix}^2$ centred in the maximum, generating a cropped version of the cross correlation: $C_5$. Simulations show that the centroid algorithms behave similarly for $C$ and $C_5$. 

The combination of the original pixel grid based conventional correlation (in a large field of view) and a sub-image grid  correlation  within a small field of view, warrants a high dynamic range shift determination to the algorithm.

\IncMargin{1em}
\begin{algorithm} \label{alg:Method2} 
    \caption{Window shift peak-finding algorithm. The function {\tt FindCentre} is one of the peak-finding algorithms presented in Tab.~\ref{tab:centroid_alg}. The function {\tt Correlation($I_\mathrm{S}; I_0$)} is given in Eq.~\ref{CCI}. The function {\tt FindCentreInteger($C$)} determines the integer pixel where the maximum of $C$ is located. The function {\tt Interpolate2D($I_{\rm S}; x; y$)} extracts an interpolated image $I_\mathrm{ROI}$ from  $I_{\rm S}$  at grid array locations $x$ and $y$.  The function  {\tt Range2D$(a,b,N)$} creates a square 2D mesh from $(a,a)$ to $(b,b)$, with $N^2$ pixels.  In Section~\ref{Methods2} the algorithm is explained in detail. The $I_{\rm ROI}$ has image dimensions as $I_0$.}
	
	\SetKwFunction{Interpolate}{Interpolate2D}
	\SetKwFunction{FindCentreInteger}{FindCentreInteger}
	\SetKwFunction{FindCentre}{FindCentre}
	\SetKwFunction{Range}{Range2D}
	\SetKwFunction{Correlation}{Correlation}
	\SetKwFunction{Average}{Average}
	
	\KwData{Reference image $I_0$; sub-aperture image $I_\mathrm{S}$; sub-sampling scale $K$.}
	\KwResult{Unbiased sub-pixel shift $s'_\mathrm{A}$.}
	
	\BlankLine
	\Begin{
		\underline{Coarse search}\;
		$C[x,y]=$ \Correlation{$I_\mathrm{S}; I_0$}\;
		$(x_0,y_0)=$ \FindCentreInteger{$C$}\;
		\underline{Fine search}\;
		\For{$k=1 $ \KwTo $K$}{ 
			$\delta = (k-1)/K$\;
			\For{$i=x_0-2 $ \KwTo $x_0+2$}{ 
				$x=$ \Range{$i - N/2 + 1 + \delta; i + N/2 + \delta; N$}\;
				\For{$j=y_0-2 $ \KwTo $y_0+2$}{ 
					$y=$ \Range{$j-N/2+1 + \delta; j + N/2 + \delta; N$}\;
					$I_{\rm ROI} = $ \Interpolate{$I_{\rm S}; x; y$}\;
					$C_5[i,j]=$ \Correlation{$I_{\rm ROI};I_0$}\;         
				} 
			} 
			$s'=$ \FindCentre{$C_5$}\;
			$s'[k] = s' -\delta$ //remove input shift applied to $I_{\rm ROI}$ at interpolation step\;
		}
		 $s'_\mathrm{A}$ $=$ \Average{$s'[k]$}\;
	}
\end{algorithm}
\DecMargin{1em}


\subsection{Synthetic sub-aperture images}\label{Synthetic}

\begin{figure*}
	\centering
	\begin{tikzpicture}
	\node[anchor=south west, inner sep=0] (image) at (0,0) 	{\includegraphics[width=0.9\textwidth]{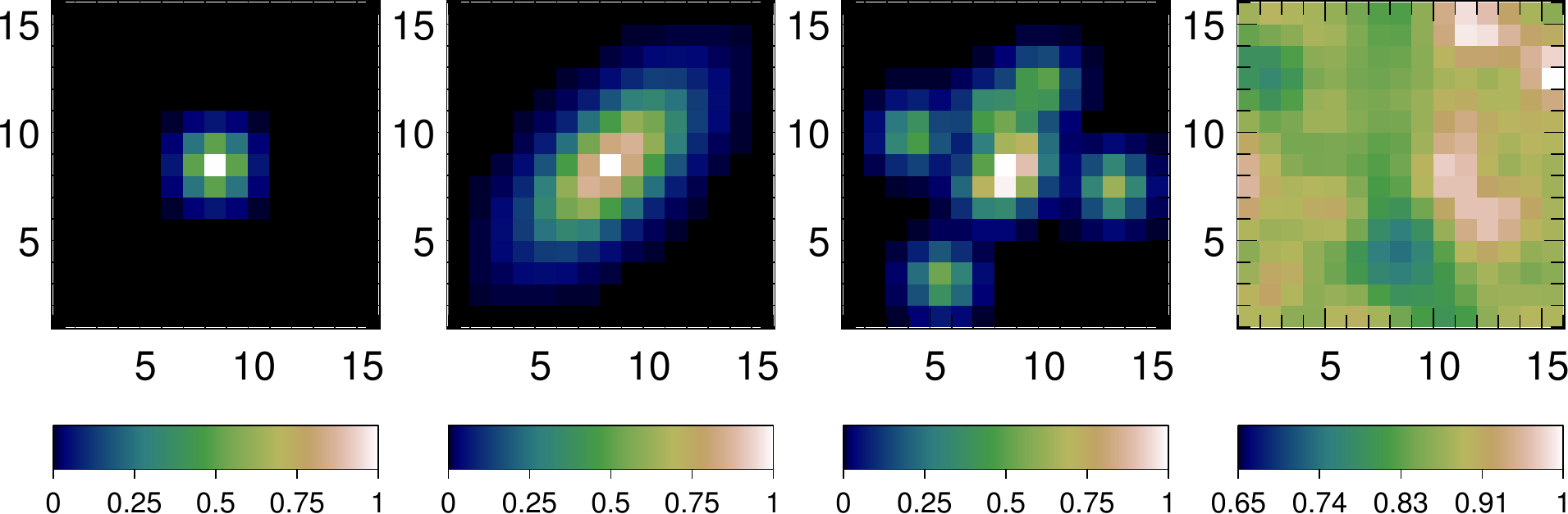}};
	\begin{scope}[x={(image.south east)},y={(image.north west)}]
	\node[left] at (image.west) 
	{\rotatebox{90}{\hspace{1cm} $j$ (pix)}};
	\end{scope}
    \node[text width=3cm] at (3.5,1.1) {$i$ (pix)};
    \node[text width=3cm] at (7.5,1.1) {$i$ (pix)};
    \node[text width=3cm] at (11.5,1.1) {$i$ (pix)};
    \node[text width=3cm] at (15.5,1.1) {$i$ (pix)};
	\end{tikzpicture}
	\caption{Synthetic Shack-Hartmann sub-aperture images. From left to right: a) the point source; b) the elongated laser guide star (LGS); c) the crowded field; d) the solar photosphere. Colorbars indicate contrast levels.}
	\label{fig:Sub-aperture-images}
\end{figure*}

Four types of sub-aperture image models of relevance for astronomical adaptive optics were used: a) a point source  diffracted spot; b) a laser guide star  elongated spot  \citep{Schreiber2009}; c) a crowded field  image; d) a solar photosphere image \citep{Lofdahl2010}. The simulation of the point source and the laser guide star are realized using 2D Gaussian profiles (circular $2 \times 2\,\mathrm{pix}^2$ and  elliptical $3 \times 6\,\mathrm{pix}^2$ with a $45^\circ$ rotation angle, respectively). The crowded field sub-aperture images are obtained by shifting and adding circular Gaussian profiles of varying intensity. To model the sub-aperture solar photosphere image, a Swedish Solar Telescope  solar granulation image is used\footnote{http://www.isf.astro.su.se/gallery/}. All the sub-aperture images are Nyquist sampled and have a  size of $16\times 16~\mathrm{pix}^2$. The synthetic sub-aperture images are presented in Fig.~\ref{fig:Sub-aperture-images}. 

The synthetic image shifts ($s$) due to atmospheric tilts are generated as follows. For the point source, laser guide star and crowed field the shifts $s$ are directly applied to the Gaussian profiles. The original solar image has a factor of 10 larger sampling than the one used for the sub-aperture images. The original image is shifted and blurred to the target Nyquist resolution by convolving it with a PSF. The resulting image is binned to generate a $16\times 16~\mathrm{pix}^2$.

Due to the extended and low contrast nature of the solar image, the cross-correlation algorithm is slightly adapted. The mean intensity is subtracted from the reference and sub-aperture solar images  because their linear intensity trend (low contrast) can shift the correlation centre from its correct position \citep{Lofdahl2010}.

Noise is added to the synthetic sub-aperture images. For all images a Gaussian read-out-noise ($\sigma_\mathrm{R}$) of $1\,\mathrm{e}^-\,\mathrm{pix}^{-1}$ is assumed,  in line with new generation detectors \citep{Finger2014, Feautrier2016}. Each synthetic image was generated with counts in each pixel following Poisson statistics. The total image signal-to-noise ratio (SNR) is calculated as 
\begin{equation}
\mathrm{SNR} = \frac{N_\mathrm{e}}{\sqrt{N_\mathrm{e} + \sigma_\mathrm{R}^2 N_\mathrm{P} }},
\end{equation}
where $N_\mathrm{e}$ and $N_\mathrm{P}$ are the total number of electrons and pixels in the sub-aperture image.
For reference, $N_\mathrm{e}=5\times 10^3\,\mathrm{e}^-$, corresponds to a 9.5~magnitude H-band star with integration time of $10^{-2}\,\mathrm{s}$, when a $9 \times 9$  lenslet and a $8~\mathrm{m}$ class telescope considered. For the solar image case $N_\mathrm{e}=5\times 10^4\,\mathrm{e}^-$ corresponds to a $\mathrm{SNR} =104$.

\section{Results and discussion}\label{RESULTS}

\subsection{Performance of current peak-finding methods}\label{centreResults}

\subsubsection{No noise case}\label{sec:fixed_snr}

\begin{figure}
	\centering
	
	\begin{tikzpicture}
	\node[anchor=south west, inner sep=0] (image) at (0,0) 	{\includegraphics[width=0.9\columnwidth]{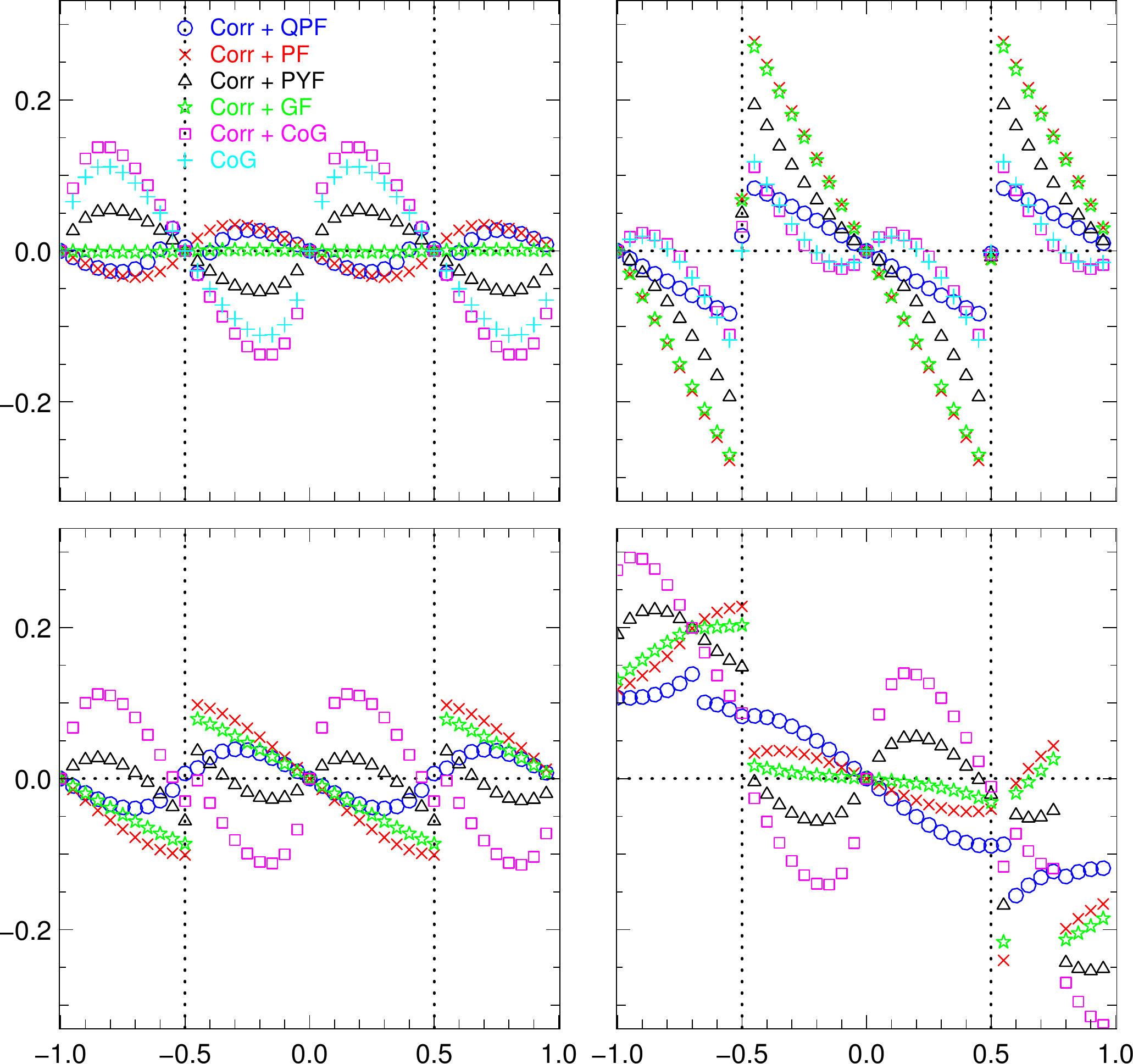}};
	\begin{scope}[x={(image.south east)},y={(image.north west)}]
	\node[below] at (image.south) 
	{\hspace{10pt}  $s$ (pix) \hspace{80pt} $s$ (pix)};
	\node[left] at (image.west) 
	{\rotatebox{90}{$\beta$ (pix)\hspace{80pt}  $\beta$ (pix)}};
	\end{scope}
	\end{tikzpicture}
	
	\caption{\label{fig:CurveFitting} Bias errors for various peak-finding algorithms and sub-aperture images. The algorithm colour/symbol legend is presented at the top left panel, cf. Table~\ref{tab:centroid_alg} for abbreviations translation. The "Corr +" label is used when the algorithm is applied to the correlation image. Sub-aperture images are: point source (top left); laser guide star (top right); crowded field (bottom left); and solar photosphere (bottom right). The shift vector is $\vec{s}=[s,s]^T$. The top left (point source) and right (laser guide star) images also include the results for direct application of the Centre of Gravity algorithm to the sub-aperture images, plus signs labelled "CoG".} 
\end{figure}

To study the bias, synthetic sub-aperture images displaced  horizontally at known positions $s$ are generated. In this section, the sub-aperture images have no noise. The positions $s$ varied from $-1\,\mathrm{pix}$ to $1\,\mathrm{pix}$, in steps of $0.05\,\mathrm{pix}$. For each synthetic image, the correlation centres $s'$ are computed with the conventional cross-correlation  as described in Sec.~\ref{sec:existing_alg}. 
Then the peak-finding algorithms of Table~\ref{tab:centroid_alg} are applied to estimate the  position $s'$. The biases are then simply $\beta=s'-s$. For completeness, the bias is also presented for the determination of $s'$ with the centre of gravity algorithm directly (i.e. without the correlation) in the sub-aperture images point source and elongated laser guide star. The results are presented in Fig.~\ref{fig:CurveFitting}, for the four types of sub-aperture images.

It is found that the bias are anti-symmetric ($\beta(-x)= -\beta(x)$) for all sub-aperture images as expected. It is periodic for the point source, laser guide star and crowded field images. But not periodic for the solar image. This non-periodicity is due to the extended scene nature of the image. When a shift is applied different parts of the image enter the field of view of the sub-aperture. Therefore the solar image does not have $\beta=0\,\mathrm{pix}$ at $s=\pm 0.5\,\mathrm{pix}$. The exact shape of the bias curve depends on the centroid algorithm and also on the nature of the image. In the $[-0.5, 0.5]\,\mathrm{pix}$ interval the bias extreme values are located approximately at $s=\pm 0.25\,\mathrm{pix}$ for all images except the laser guide star\footnote{The shape of the $\beta$ curve for the laser guide star is due to applying shift $\vec{s}=[s,s]^T$ along the diagonal of the image. For a horizontal shift, it would have a similar shape as the point source and crowded field.}.

Sharp transitions are observed for extended sub-aperture images at pixel positions $s=\pm 0.5\,\mathrm{pix}$. This is due to the shift vector being diagonal:  $\vec{s}=[s,s]^T$, which translates in the peak of the correlation being "split" into two diagonal pixels. When the shift $s$ is, e.g. $0.4\,\mathrm{pix}$, the brighter pixel is in the lower left one and the bias is negative. When the shift is e.g.  $0.6\,\mathrm{pix}$, the brighter pixel is  the top right and the bias is positive. 

The best performing centroid algorithm depends on the image: a) for the point source and crowded field, it is the Gaussian fit; b) for the solar image it is 2D quadratic polynomial; c) for the laser guide star it is the centre of gravity. The reason for this behaviour is the matching of the algorithm to the actual shape of each image  correlation centre (e.g. point source and crowded field are generated with circular Gaussians).

\subsubsection{Varying signal-to-noise ratio}\label{sec:varying_snr}

\begin{figure*}
	\centering
	
	
	\parbox[t]{11pt}{\rotatebox{90}{\hspace{0.12\textwidth}  $\sigma$ (pix)}}
	\includegraphics[width=0.28\textwidth]{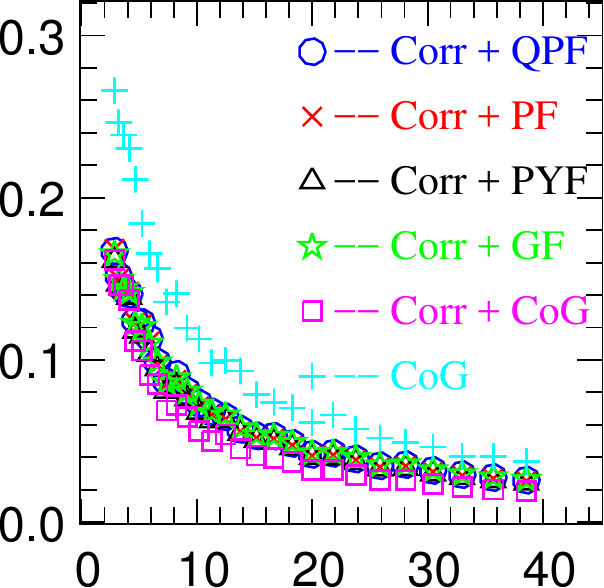} 
	\hspace{5pt} 
	\parbox[t]{11pt}{\rotatebox{90}
		{\hspace{1.6 cm} $|\left<s'\right>-s\,|$~~(pix)}}
	\includegraphics[width=0.28\textwidth]{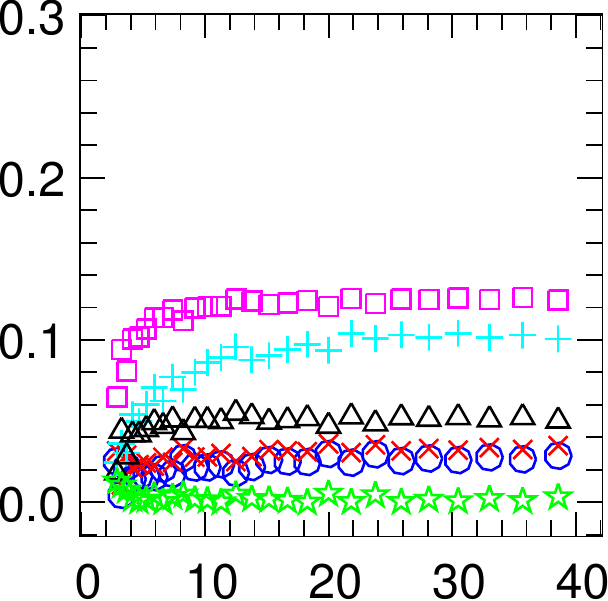}  
	\hspace{5pt} 
	\parbox[t]{11pt}{\rotatebox{90}
		{\hspace{1.6 cm} $|\left<s'\right>-s\,|$~~(pix)}}
	\includegraphics[width=0.28\textwidth]{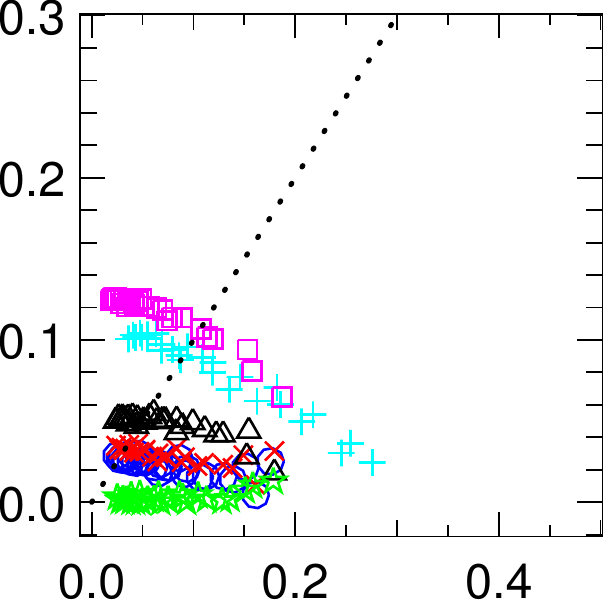} 
	
	\vspace{11pt}
	
	\parbox[t]{11pt}{\rotatebox{90}{\hspace{0.12\textwidth}  $\sigma$ (pix)}}
	\includegraphics[width=0.28\textwidth]{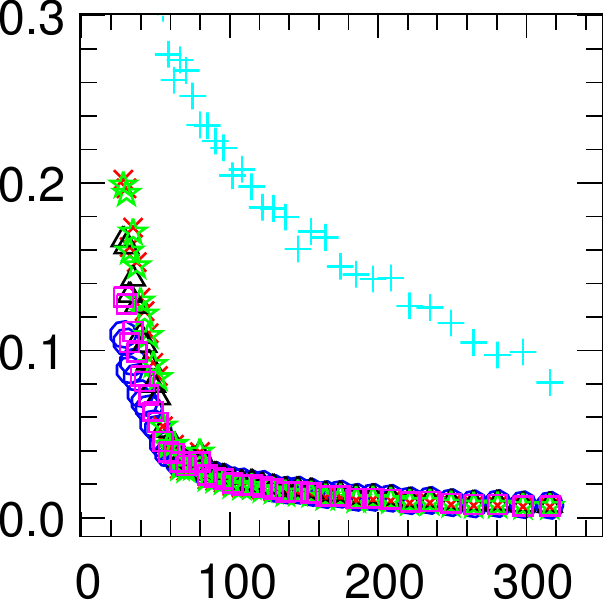}
	\hspace{5pt}
	\parbox[t]{11pt}{\rotatebox{90}
		{\hspace{1.6 cm} $|\left<s'\right>-s\,|$~~(pix)}}
	\includegraphics[width=0.28\textwidth]{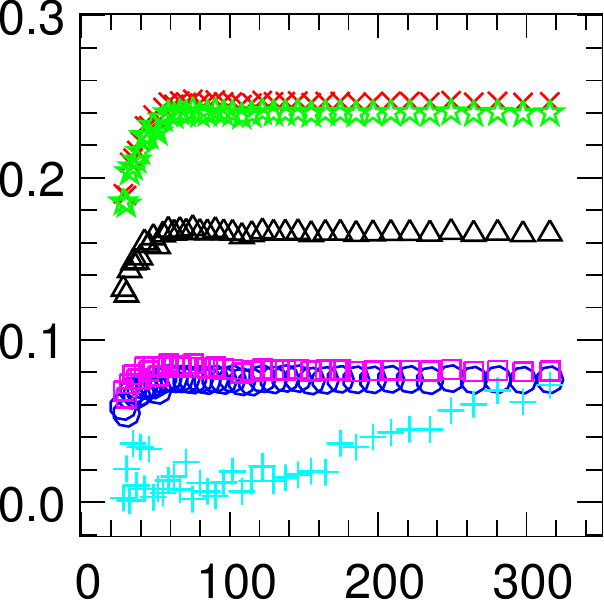} 
	\hspace{5pt}
	\parbox[t]{11pt}{\rotatebox{90}
		{\hspace{1.6 cm} $|\left<s'\right>-s\,|$~~(pix)}}
	\includegraphics[width=0.28\textwidth]{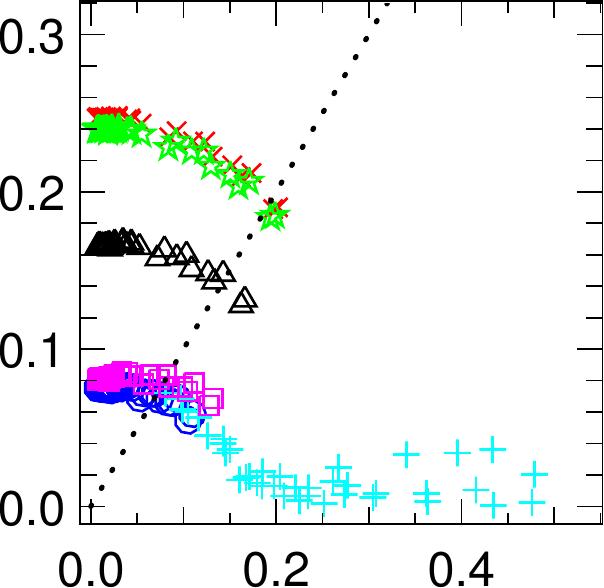} 
	
    \vspace{11pt}
	
	\parbox[t]{11pt}{\rotatebox{90}{\hspace{0.12\textwidth}  $\sigma$ (pix)}}
	\includegraphics[width=0.28\textwidth]{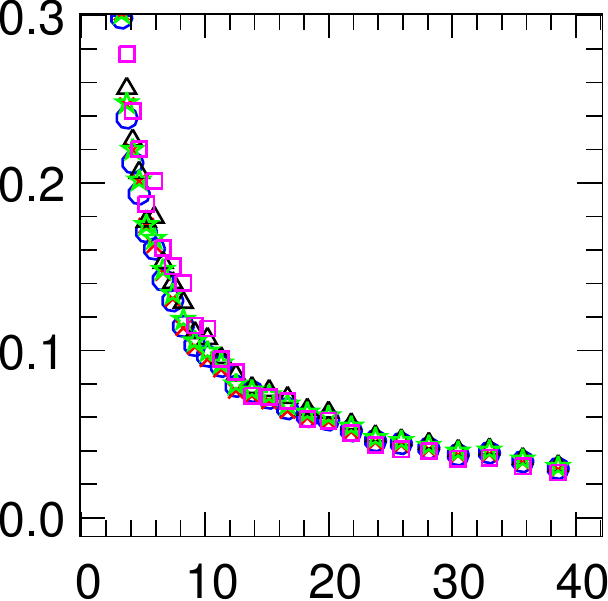}
	\hspace{5pt}
	\parbox[t]{11pt}{\rotatebox{90}
		{\hspace{1.6 cm} $|\left<s'\right>-s\,|$~~(pix)}}
	\includegraphics[width=0.28\textwidth]{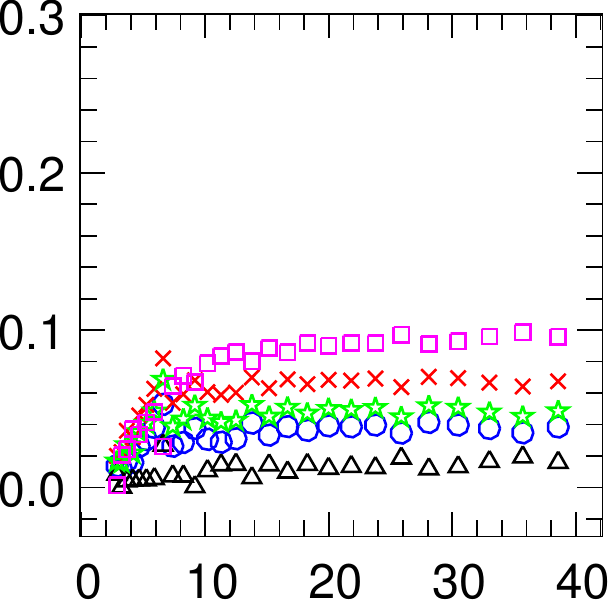} 
	\hspace{5pt}
	\parbox[t]{11pt}{\rotatebox{90}
		{\hspace{1.6 cm} $|\left<s'\right>-s\,|$~~(pix)}}
	\includegraphics[width=0.28\textwidth]{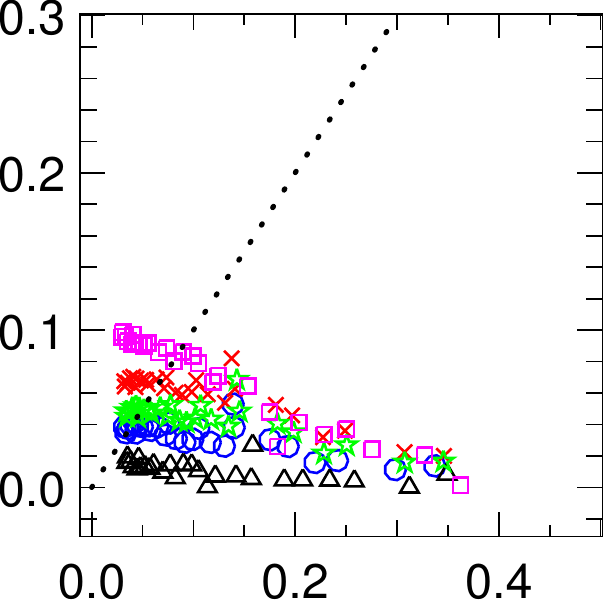}
	
	\vspace{11pt}
	
	\parbox[t]{11pt}{\rotatebox{90}{\hspace{0.12\textwidth}  $\sigma$ (pix)}}
	\includegraphics[width=0.28\textwidth]{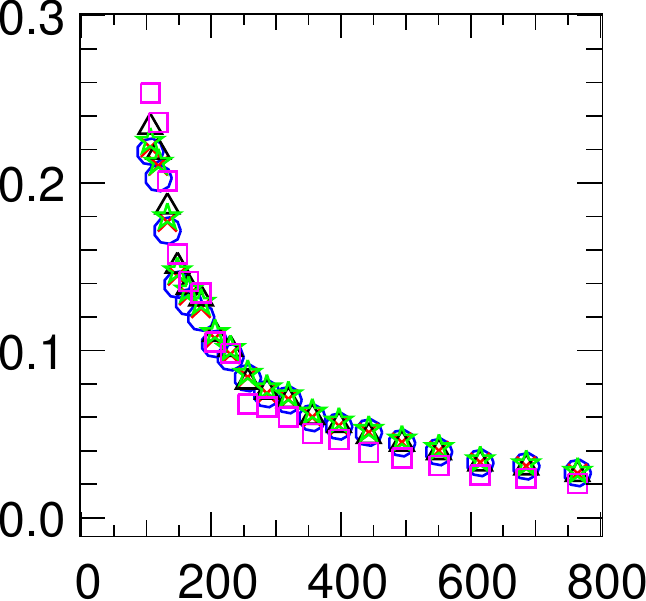} 
	\hspace{5pt}
	\parbox[t]{11pt}{\rotatebox{90}
		{\hspace{1.6 cm} $|\left<s'\right>-s\,|$~~(pix)}}
	\includegraphics[width=0.28\textwidth]{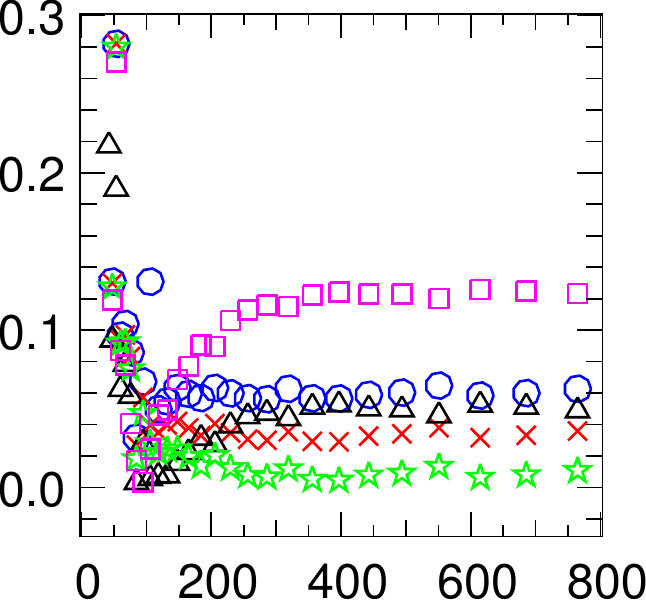}  
	\hspace{5pt}
	\parbox[t]{11pt}{\rotatebox{90}
		{\hspace{1.6 cm} $|\left<s'\right>-s\,|$~~(pix)}}
	\includegraphics[width=0.28\textwidth]{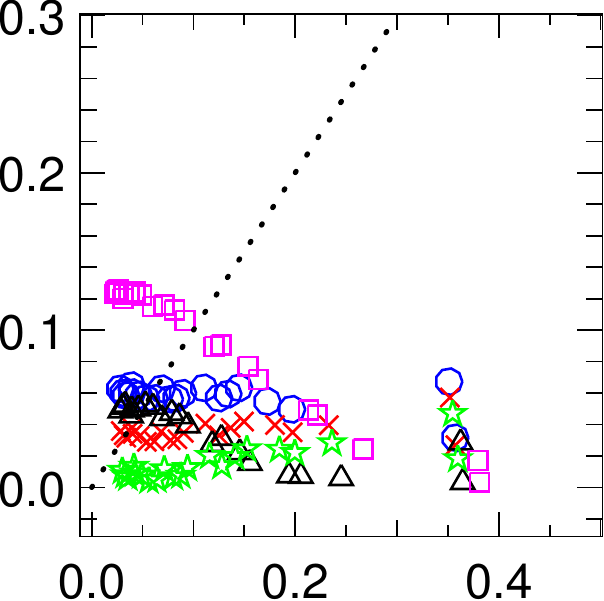} 
	\\[5pt]
	{ \hspace{1cm} SNR \hspace{5cm} SNR \hspace{4.5cm} $\sigma$ (pix)}
	
	\caption{\label{fig:SNR}  Performance of centroid algorithms with SNR. Left column: RMS centroid error $\sigma$; centre column: $|\left<s'\right>-s|$, which coverages to bias error $\beta$ at high SNR; right column $\left<s'\right>-s$ versus $\sigma$. The dotted line in the right column traces $\left<s'\right>-s= \sigma$.   The rows from top to bottom are: point source, laser guide star, crowded field, solar image, respectively. The dotted line in the last column is where bias error and random error is equal. The data points above this dotted line indicate bias error domination over random error. }
	
\end{figure*}

The centroid algorithms' performance was tested in varying SNR conditions.  For each SNR, $500$ random realizations are generated at an input shift vector $\vec{s}=[s_\mathrm{max},s_\mathrm{max}]^T$ where the bias  is approximately maximum: a) $s_\mathrm{max}=0.25\,\mathrm{pix}$ for the point, crowded field and solar images; b)  $s_\mathrm{max}=0.4\,\mathrm{pix}$ for the laser guide star image. Note that the exact maximum location depends on the algorithm and therefore $s_\mathrm{max}$ is approximate. The input shifts are measured by applying the conventional cross-correlation  (Eq.~\ref{CCI}) and centroid algorithm (cf. Table~\ref{tab:centroid_alg}). The bias  and SNR are computed using Eq.~\ref{eq:BiasError} and Eq.~\ref{SNR}, respectively. The SNR is varied via $N_\mathrm{P}$ in the sub-aperture images. 

The results are presented in Fig.~\ref{fig:SNR}.  The first column presents the RMS centroid error ($\sigma$). It is computed by the RMS of $s'$. It decreases  with SNR as expected. The direct image centre determination via the Centre of Gravity algorithm has a worse behaviour for the point source and laser guide star than the correlation algorithms. The superiority of the correlation with respect to $\sigma$ is well known in the literature \citep[e.g.][]{Thomas2006}. Intuitively it is expected because of the noise smoothing and shape matching. All correlation algorithms have a similar behaviour.

The centre column of Fig.~\ref{fig:SNR} shows that $|\left<s'\right>-s\,|$  increases to an asymptotic value, the bias error $\beta$. $|\left<s'\right>-s\,|$ is not constant because at low SNR the effective image shape changes. Intuitively the low value of $|\left<s'\right>-s\,|$ at low SNR can be explained by a pure noise image, for which the bias is expected to be zero. The large values of $|\left<s'\right>-s\,|$ for the low SNR of the solar image (centre column, bottom row of Fig.~\ref{fig:SNR}) are due to the large variance of  $|\left<s'\right>-s\,|$ for this sub-aperture image. The large bias for the laser guide star in comparison to others is  caused by the shape of the correlation peak, which is elongated and oriented $45^\circ$ rotation angle.   

The right column of Fig.~\ref{fig:SNR} plots $|\left<s'\right>-s\,|$ versus $\sigma$. The dashed line is the imaginary curve $|\left<s'\right>-s\,|=\sigma$. Points above the curve show a bias error larger than the centroid error. Typically the bias error is larger than the noise error for SNR larger than 10, except for the solar case where it becomes important for SNR larger than 200. 

\subsection{Performance of the window shift method}\label{TwoStepResults1}

In this section the performance of the algorithm introduced in Section~\ref{Methods2} is presented, initially for a fixed SNR and then for a varying SNR. 

\subsubsection{Fixed signal-to-noise ratio}\label{sec:two-step_snr_fixed}

\begin{figure}
	\centering
	
	\begin{tikzpicture}
	\node[anchor=south west, inner sep=0] (image) at (0,0) 	{
		\includegraphics[width=0.9\columnwidth]{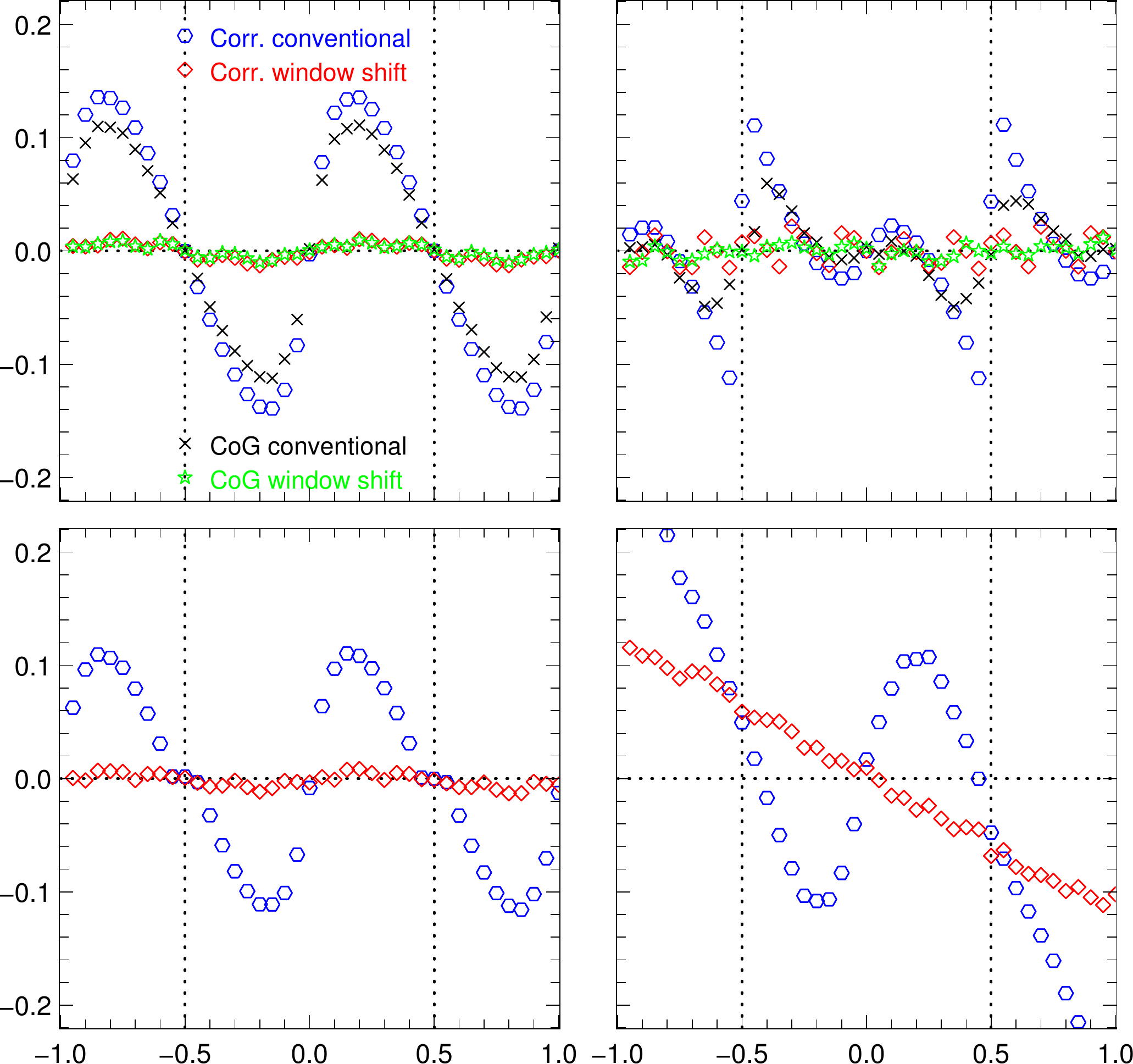}};
	\begin{scope}[x={(image.south east)},y={(image.north west)}]
	\node[below] at (image.south) 
	{\hspace{1mm} $s$ (pix)\hspace{65pt} $s$ (pix)};
	\node[left] at (image.west) 
	{\rotatebox{90}{\hspace{1mm} $\beta$ (pix) \hspace{3cm}  $\beta$ (pix)}};
	\end{scope}
	\end{tikzpicture}
	
	\caption{\label{fig:CorrelationPerformance} Bias errors of the window shift method in comparison with the conventional approach. The algorithm legend is presented at the top left panel. Sub-aperture images are: point source (top left); laser guide star (top right); crowded field (bottom left); and solar photosphere (bottom right).  The shift vector is $\vec{s}=[s,s]^T$. The top row also includes the performance for direct sub-aperture images (labelled CoG).}
\end{figure}

Fig.~\ref{fig:CorrelationPerformance} shows the comparison of the window shift method with the conventional cross-correlation algorithm. One of the worst performing centroid algorithms -- the centre of gravity -- was used. The SNR conditions are the same as in Section~\ref{sec:fixed_snr}. The sampling factor is $K=5$. The window shift method drastically reduces the bias. For the solar image the final bias is larger, because the shift will include pixels from the edge of the image that are not present in the reference image.

\subsubsection{Varying sampling factor $K$}

\begin{figure}
	\centering
	
	\parbox[t]{11pt}{\rotatebox{90}{\hspace{2.2cm}  $\beta$ (pix)}}
	\includegraphics[width=0.7\columnwidth]{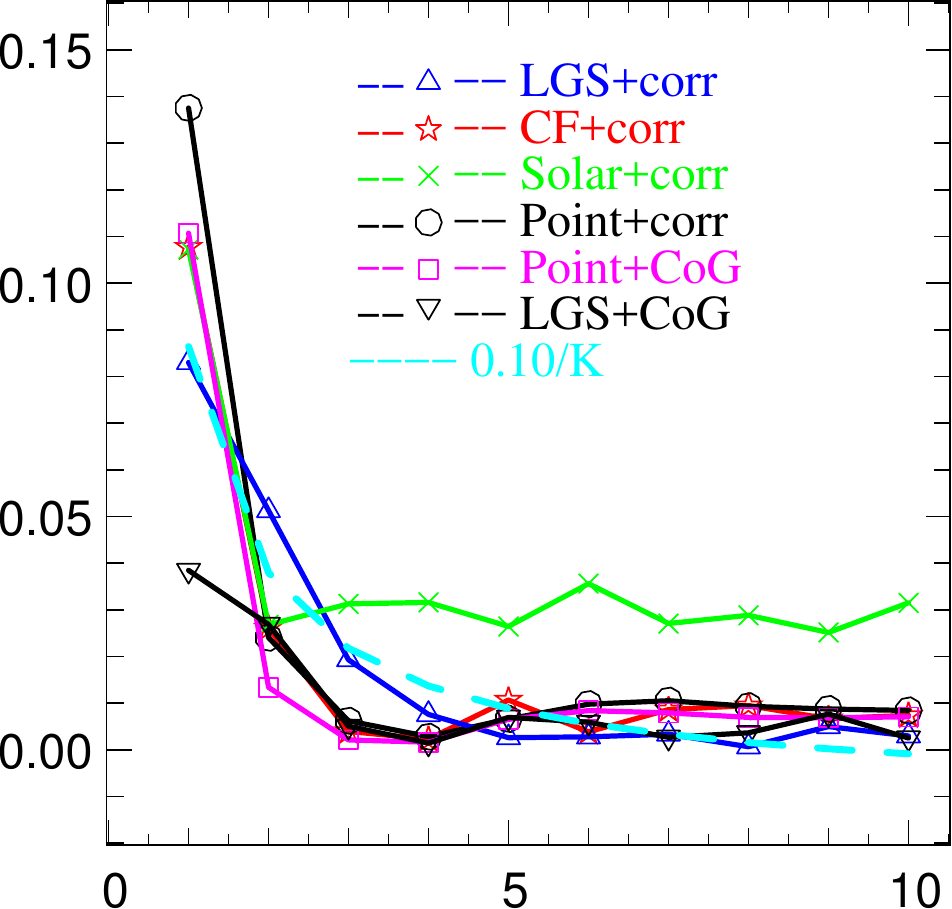} 
	
	\hspace{16pt}  $K$ 
	
	\caption{\label{fig:Resolution} Performance of the window shift method as a function sampling factor $K$. The curves are for: point source, laser guide star (LGS), crowded field (CF) and solar image. The dashed curve depicts the function $\beta(K)=0.1/K$.}
\end{figure}

The effect of the sampling factor $K$ in reducing the bias $\beta$ of the centroid algorithm is presented in Fig.\ref{fig:Resolution}. The setup is the same as Section~\ref{sec:two-step_snr_fixed}, except for the sampling factor $K$, which varied. The centre of gravity algorithm was used considering its better performance against lower SNR for point, crowded field and solar images.

The bias $\beta$ strongly decreases with the sampling factor $K$. It approximately follows a $\propto K^{-1}$ relation. For sampling factors $K>5-6$ no significant improvement is observed. This behaviour is similar to the one observed by \cite{Gui2002} in a different context.

\subsubsection{Varying signal-to-noise ratio}

\begin{figure}
	\centering
	\parbox[t]{11pt}{\rotatebox{90}{\hspace{1.2cm}  $\beta$ (pix)}}
	\includegraphics[width=0.9\columnwidth]{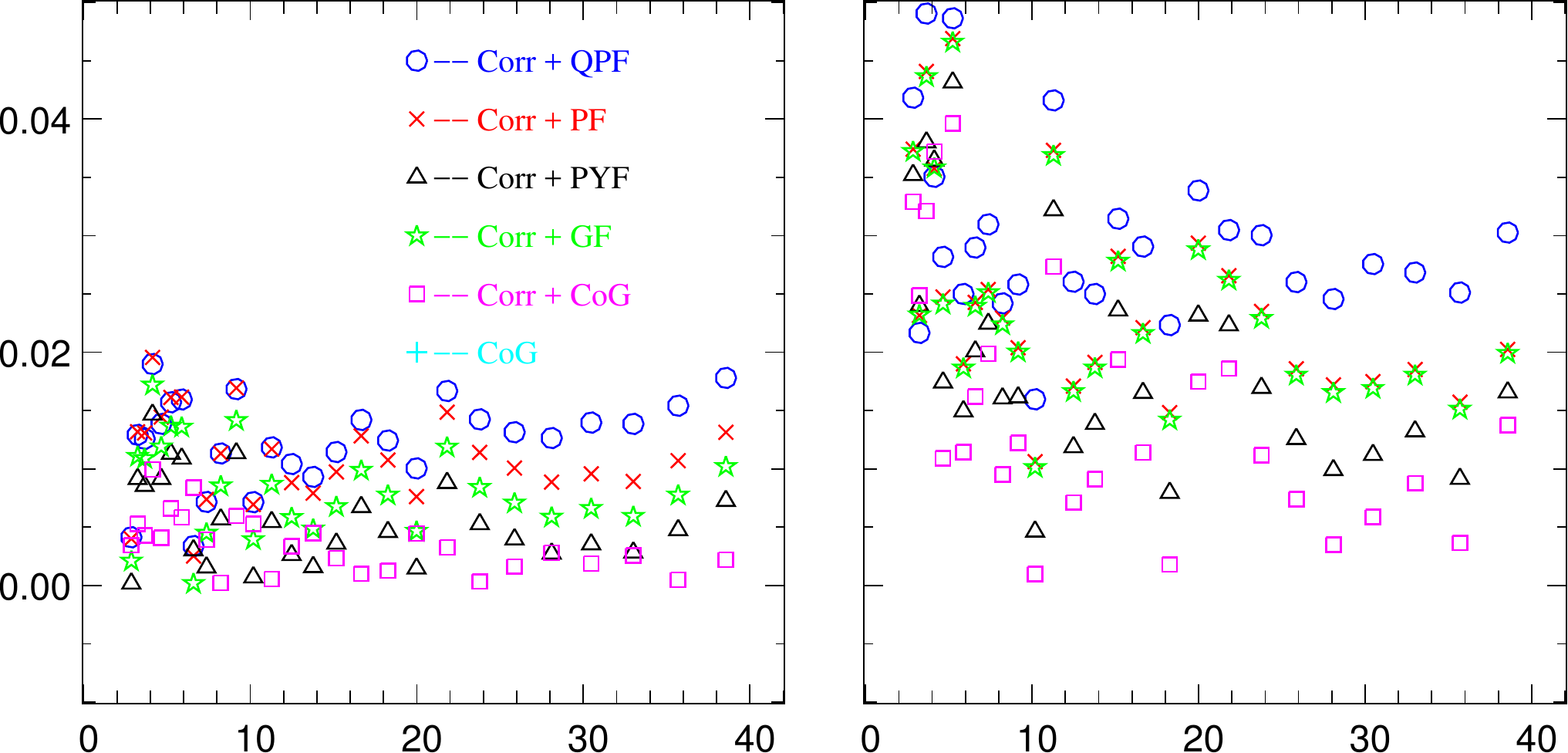} 
	\hspace{0.5\columnwidth} SNR \hspace{0.35\columnwidth} SNR
	
	\vspace{5pt}
	
	\parbox[t]{11pt}{\rotatebox{90}{\hspace{1.2cm}  $\beta$ (pix)}}
	\includegraphics[width=0.9\columnwidth]{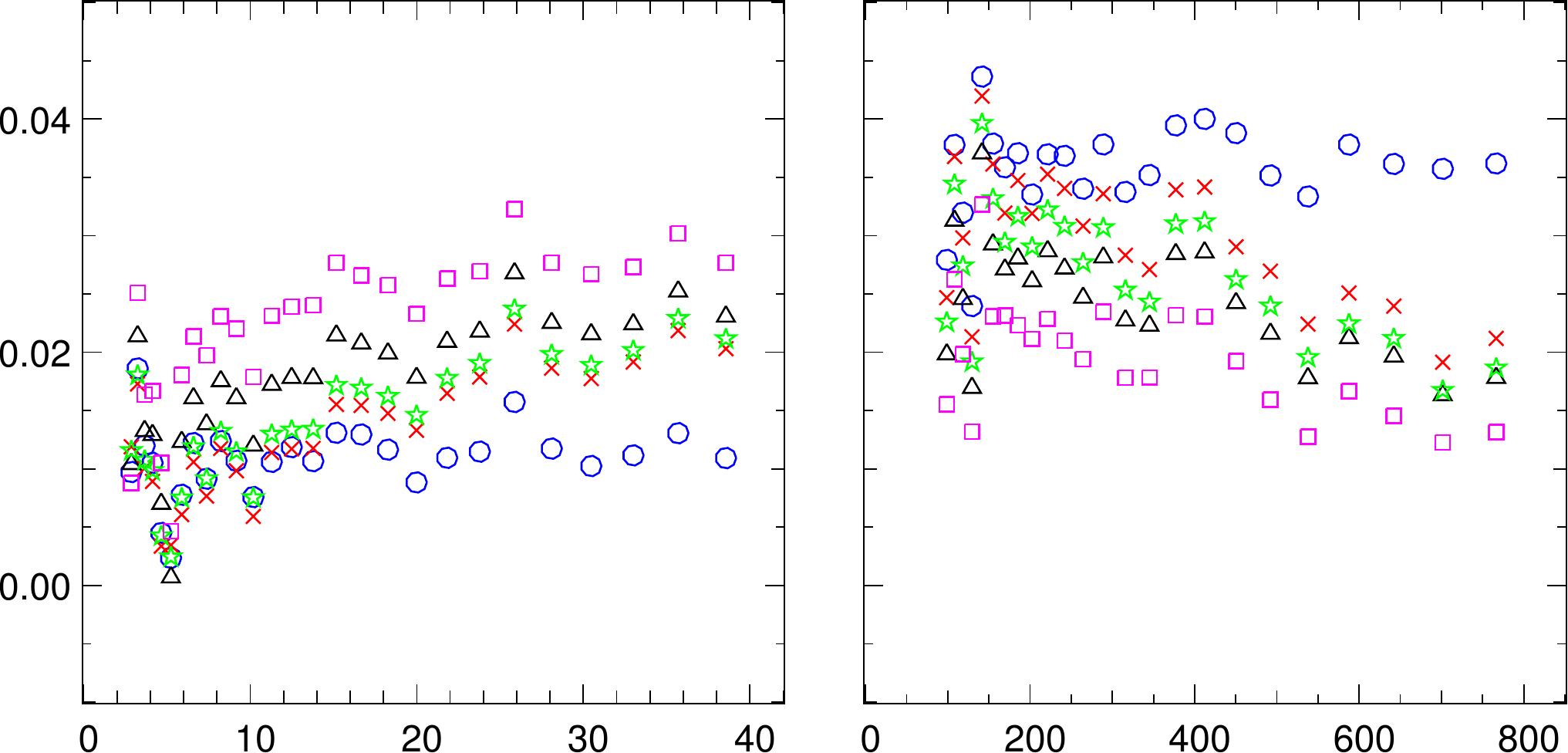}
	\hspace{0.5\columnwidth} SNR \hspace{0.35\columnwidth} SNR
	
	\caption{\label{fig:SNR_TwoStep}  Residual bias errors for window shift method as a function of SNR: point source (top left); laser guide star (top right); crowded field (bottom left); and solar photosphere (bottom right). } 
\end{figure}

The performance of the window shift method  as a function of SNR is presented in Fig.~\ref{fig:SNR_TwoStep}. The same setup as the one presented in Section~\ref{sec:varying_snr} is used (e.g. $s$ position). The sampling factor is $K=5$. The  residual errors are well below 0.05 pix for all SNR and sub-aperture image type.

The performance is similar to the windowed, adaptive thresholding centre of the mass method of \cite{Townson2015}\footnote{Note that a different SNR metric is used in \cite{Townson2015}, i.e. in their Figure~7 a SNR of 20 corresponds to an $\mathrm{SNR}\sim 100$ in our Figure~\ref{fig:SNR_TwoStep}.}.

\subsubsection{Computational efficiency}

Several number of factors can influence the total execution time of the algorithm, importantly, the sub-aperture window size, the efficiency of programming (ex. multithreading), the performance of hardware and the programming language. Therefore only the relative computational efficiency was computed.  The proposed window shift method is  slower in comparison to the conventional algorithm by a factor of 2.4 and 3.6 for $K=3$ and $K=5$, respectively. For  larger sub-apertures, the computational time ratio is reduced because the window shift method uses a fixed and small correlation sub-image $C_5$.  If the sub-aperture size is increased from  $16 \times 16\,\mathrm{pix}^2$ to $32 \times 32\,\mathrm{pix}^2$  the window shift method is slower by a factor of  $1.4$ and $1.8$ for $K=3$ and $K=5$, respectively.

\section{Conclusions}\label{sec:conclusion}

A systematic study of the bias error in conventional centroid algorithms used for slope measurement in Shack-Hartmann wave-front sensors is presented for the first time. It is found that the bias can be as the same order of magnitude of the centroid error, especially at moderate and high SNR ratios, typically the bias error is larger than the noise error for SNR larger than 10, except for the solar case where it becomes important for SNR larger than 200.

No centroid method reduces both the bias and noise error terms  in conventional correlation methods.  A window shift method is proposed based on the anti-symmetric nature of the bias. It works by sampling the sub-aperture image $K$ times, at the same resolution, but shifted by a sub-pixel step, with size function of $K$. The obtained $K$ shifts are then averaged out,  significantly  reducing the bias. The window shift method is studied as a  function of image type, centroid algorithm, SNR and $K$ sampling factor. It is found that it robustly reduces the bias by a factor of $\sim 7$ to values of $\lesssim 0.02~\mathrm{pix}$. The computational cost of the algorithm is optimized by obtaining the correlation in two steps: a) large region based coarse search; b) small region  based $C_5$ fine search.  It ranges between a factor of 1.4  to 3.6 of conventional approaches.

The window shift method can be applied to other algorithms which work similar to the cross-correlation algorithm such as square difference function, absolute difference function and square of the absolute difference function \citep{Lofdahl2010}. The square difference function is especially important for the solar type of images as it gives a significantly smaller random error and more anti-symmetric pattern of systematic error~\citep{Lofdahl2010}. However, the systematic error values are larger in a comparison to the cross-correlation. The proposed method would be of relevance for the square difference function to reduce its systematic error by using its consistent anti-symmetric pattern. 

Further developments are the study of the window shift algorithm for sub-aperture images that have a sampling smaller than the critical sampling and for Shack-Hartmann devices with a small number of apertures, such as those used for fast tip-tilt correction or pupil tracking.

\section*{Acknowledgements}
	
We are grateful to  Lancelot, J. P and Akondi, V  for commenting and proof reading of the manuscript. We thank the referee for her/his constructive and insightful report. This research was partially supported by Funda\c{c}\~{a}o para a Ci\^{e}ncia e a Tecnologia (contracts PTDC/CTE-AST/116561/2010, SFRH/BD/52066/2012, UID/FIS/00099/2013) and the European Commission (EC) (Grant Agreement 312430). 
MC wishes to acknowledge A*MIDEX project (no. ANR-11-IDEX-0001-02) funded by the French Government programme "Investissements d'Avenir", managed by the French National Research Agency (ANR).









\bsp	
\label{lastpage}
\end{document}